\documentclass[final,3p,times]{elsarticle}

\usepackage{amssymb}
\usepackage{amsmath}
\usepackage{amsthm}

\journal{Chaos, Solitons and Fractals}
\begin{document}
	
	\begin{frontmatter}
		
		
		
	\title{Atypical Chimera States in an Ensemble of Partially Mobile Particles}
		
		
		
		
		\author[aff1]{Pavel A. Shcherbakov\corref{cor1}}
		\cortext[cor1]{Corresponding author}
		\ead{pavel.scherbakov@unn.ru}
		\author[aff1]{Lev A. Smirnov}
		\author[aff2,aff1]{Vasily~A.~Kostin}
		\author[aff1]{Maxim I. Bolotov}
		\author[aff1]{Grigory V. Osipov}
		
		\affiliation[aff1]{organization={Department of Control Theory, Lobachevsky State University of Nizhny Novgorod},
			addressline={Gagarin~Ave.~23}, 
			city={Nizhny Novgorod},
			postcode={603022}, 
			country={Russia}}
		
		\affiliation[aff2]{organization={Gaponov-Grekhov Institute of Applied Physics of the Russian Academy of Sciences},
			addressline={Ul’yanova Str.~46}, 
			city={Nizhny Novgorod},
			postcode={603950}, 
			country={Russia}}
		
%
%
%
%
		\begin{abstract}
			We study the influence of nonuniform motion of oscillators in a ring chain with nonlocal coupling on their collective dynamics and reveal the mechanism behind the emergence of an atypical chimera state in such systems. The mechanism relies on regular spatially inhomogeneous motion of oscillators, which breaks the symmetry of the effective interaction kernel. This symmetry breaking induces spatial phase correlations in the asynchronous part of the system, giving rise to nonuniformly twisted and previously unobserved coherent–incoherent–twisted states.
		\end{abstract}
		
		
		
		
	\end{frontmatter}
	
		
		
		\section{Introduction}
		\label{sec1}
		
		Synchronization underlies the collective behavior of numerous natural and engineered systems comprising large ensembles of interacting units \cite{pikovsky2001synchronization, Vicsek2012, hong2005collective}. A widely adopted framework for analyzing such systems is that of coupled phase oscillators. When interactions are sufficiently weak, the dynamics of each unit can be reduced to a single phase variable \cite{hoppensteadt2012weakly, kuramoto84, PhysRevE.100.012211}. Depending on the coupling topology, both global and nonlocal coupling schemes are commonly studied \cite{RevModPhys.77.137, PhysRevE.82.016205}. In systems with nonlocal coupling, one of the characteristic dynamical regimes is the chimera state, where spatially localized clusters of synchronized oscillators coexist with regions of incoherent dynamics \cite{kuramoto2002coexistence, PhysRevLett.93.174102}. Such states result from spontaneous symmetry breaking, as even in ensembles of identical oscillators with purely attractive coupling, the fully synchronized state may coexist with other partially coherent patterns \cite{motter2010spontaneous, panaggio2015chimeracoex}.
		
		Both stationary chimera states, in which the degree of local synchronization at each spatial point remains constant in time \cite{smirnov2017chimera}, and various nonstationary variants have been reported. These include breathing \cite{zhu2023self,suda2020emergence,omel2022mathematical}, traveling \cite{xie2014multicluster}, and alternating chimera states, in which the synchronized domain periodically disappears and reappears at different spatial locations \cite{haugland2015self, PhysRevE.91.022817}.
		
		A complementary line of research in nonlinear dynamics focuses on systems of mobile agents \cite{Bechinger2016}. While stochastic descriptions such as Brownian motion are possible \cite{Reimann2002,Hanggi2009}, deterministic models become particularly relevant in the superactive limit \cite{Lowen2020}. Active particle models have been successfully employed, for example, to describe the motion of colloidal particles in external potentials \cite{Bechinger2016}. When individual agents possess intrinsic dynamics that influence their collective behavior, their state can be effectively represented by a phase variable, as exemplified by swarmalator-based models of chemotaxis~\cite{tanaka2007general}. Such systems may be viewed as ensembles of coupled phase oscillators with moving elements \cite{Smirnov2021, bolotov2025chimera}.
		In the context of chimera states, particle motion can either separate the system into multiple coexisting chimeras~\cite{bolotov2025chimera} or promote the dominance of chimera patterns~\cite{Smirnov2021}.

		In this work, we examine how regular spatially inhomogeneous motion of phase oscillators arranged in a ring configuration with nonlocal coupling shapes possible chimera states. 
		Such kinematics can correspond to agents with mobility dependent on the spatial coordinate: under the action of an external coherent forcing field, particles—localized around some point—are involved in strong (high-amplitude) motion, while the rest move only slightly or remain still.

		Section~\ref{model} introduces the model and describes the imposed periodic motion of the oscillators. Section~\ref{fs_pm} investigates how oscillator mobility affects the bistability between the fully synchronized state and the chimera state, leading to the emergence of a periodically traveling chimera. Section~\ref{altern} demonstrates the formation of an alternating chimera state and shows that the displacement of the synchronized cluster is caused by an asymmetry in the effective nonlocal interaction kernel. By analyzing phase correlations on the set of moving particles, we further interpret the chimera as either a nonuniformly twisted state, in which the phase profile is continuous, or a coherent-incoherent-twisted state, in which the phase profile possesses a localized disruption. The main findings are summarized in the Conclusion.
		
		\section{Model}
		\label{model}
		
		We consider a system of $N$ oscillators located on a ring of length $L$. The oscillators positions are determined by coordinates $x_n \in [0,L)$ (index $n = 1,2,\ldots,N$). The internal state of each oscillator is described by the respective phase variable $\varphi_n$. The positions are time-dependent, $x_n = x_n(t)$. Particle coordinates evolve according to
		\begin{equation}
			\label{eq_A}
			x_n(t) = x_n^{(0)}+A_n \sin{\Omega t}, 
		\end{equation}
		where $x_n^{(0)} = nL/N$. We gradually introduce nonzero amplitudes of motion by localizing elements with nonzero amplitudes in a subsegment of the ring and controlling their number via an additional parameter $\sigma$, the half-width of the subsegment.   Because the medium is periodic, without loss of generality we can position our subsegment symmetrically around the point $x = L/2$. The case of the stationary equidistant particle grid corresponds to $\sigma = 0$ and has been comprehensively treated in~\cite{kuramoto2002coexistence, smirnov2017chimera}, while the stationary nonequidistant particle grid was considered in~\cite{Smirnov2021}.
		We focus on the case in which nonzero amplitudes follow a uniform probability distribution on a segment $[\mu-\Delta, \mu+\Delta]$, where $\Delta$ is the half-width and $\mu$ is the mean value. Let $\zeta_n$ be the $n$th sample drawn from this distribution, yielding  amplitudes of motion
		\begin{equation}
			\label{eq4}
			A_n =\begin{cases}
				\zeta_n&\text{if }\lvert x_n^{(0)}-L/2\rvert \leq \sigma, \\
				0&\text{otherwise.}
			\end{cases}
		\end{equation}

		The phases follow the dynamics of the classical model of Kuramoto and Battogtokh~\cite{kuramoto2002coexistence}
		\begin{equation}
			\label{eq1}
			\dot{\varphi}_n = \frac{1}{N}\sum_{j = 1}^N G\left(x_j(t)-x_n(t)\right)\sin{(\varphi_j-\varphi_n-\alpha)}, 
		\end{equation}
		where $\alpha$ is the phase shift that defines the phase interaction. The coupling between oscillators is nonlocal and is governed by the coupling kernel~\cite{smirnov2017chimera}
		\begin{equation}
			\label{eq2}
			G(x) = \kappa \cosh{\bigl(\kappa\left\lvert\{x/L\}-1/2\right\rvert\bigr)}\bigl/ 2\sinh{\bigl(\kappa L /2\bigr)}, 
		\end{equation}
		where the operation $\{x/L\} = x/L - \lfloor x/L\rfloor$ denotes the remainder when $x$ is divided by $L$, and the parameter $\kappa$ determines the strength of the nonlocal coupling. In the limit $\kappa L \to + \infty$, the expression in the Eq.~\eqref{eq2} approaches the exponential kernel $G_{\mathrm{KB}}(x) = \kappa \exp{\bigl(-\kappa\lvert x\rvert/2\bigr)}$, initially considered by Kuramoto and Battogtokh~\cite{kuramoto2002coexistence}. Both $G_{\mathrm{KB}}(x)$ and $G(x)$ are Green's functions of the inhomogeneous Helmholtz equation on an infinite line and on a ring respectively.

		To describe the degree of oscillators phase coherence in the vicinity of every oscillator we consider the complex field
		\begin{equation*}
			H_n(t) = \frac{1}{N} \sum_{j=1}^N G(x_j-x_n)\exp{i\varphi_j}.
		\end{equation*}
		To quantify the overall degree of synchronization we evaluate the global order parameter
		\begin{equation*}
			R(t) = \frac{1}{N} \sum_{j=1}^N \exp{i\varphi_j}.
		\end{equation*}
		 Here, \( \lvert R \rvert = 1 \) indicates complete synchronization, \( 0 < \lvert R\rvert < 1 \) partial synchronization, and \( R = 0 \) an incoherent state.
		We additionally define $M$ to be the set of indices $m$ of moving particles: $M = \{\,m \mid \lvert x_m^{(0)}-L/2\rvert \leq \sigma\,\}$.

		 \section{Fully synchronous state. Periodically moving chimera state}
		 \label{fs_pm}
		 For direct numerical integration of Eqs.~\eqref{eq_A}--\eqref{eq2}, we adopt the fourth-order Runge--Kutta method with a constant time step $\mathrm{d}t = 0.01$. 
		 Let us now describe the initial conditions and fixed parameters employed in our simulations. Initial phases $\varphi_n(0)$ are uniformly distributed on $(-\pi, \pi]$. The system size is fixed at $N = 512$. In our numerical simulations, we use the classical parameter values considered by Kuramoto and Battogtokh in \cite{kuramoto2002coexistence}: coupling strength $\kappa = 4.0$, length of the medium $L = 1.0$, phase shift $\alpha = 1.457$; this set of parameters allows us to obtain a stationary chimera state. We consider slow particle motion and fix the low oscillation frequency $\Omega = 0.01$ to avoid locking with the stationary chimera state that arises on the equidistant grid in the limit $N \gg 1$ (see \cite{smirnov2017chimera}). The classical model of Kuramoto and Battogtokh possesses bistability: for random initial phases, either the fully synchronous regime or the chimera state emerges with certain probabilities~\cite{smirnov2017chimera}.
			
\begin{figure}
			\includegraphics[width = 1.0\linewidth]{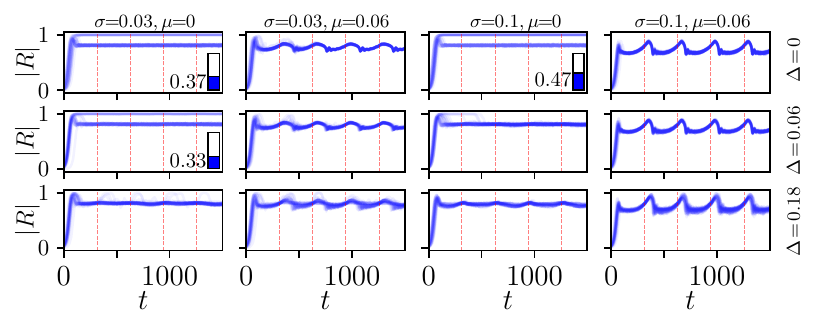}
			\caption{Dynamics of the modulus of the global order parameter $|R(t)|$ for $30$ samples of random initial conditions $\varphi_n(0)$, uniformly distributed on $(-\pi,\pi]$, for a few samples of parameter sets $(\sigma, \mu, \Delta)$. Temporal evolutions with $\lvert R(t)\rvert = 1$ correspond to the fully synchronized state. Temporal evolutions with $\lvert R(t)\rvert < 1$ correspond to chimera-like regimes. Vertical red dashed lines indicate the time instants at which the particle coordinates $x_n(t)$ are equidistantly spaced. Parameters $\sigma = 0.03$ (left two columns), $\sigma = 0.1$ (right two columns), $\mu = 0$ (first and third columns), $\mu = 0.06$ (second and fourth columns), $\Delta = 0$ (first row), $\Delta = 0.06$ (second row), $\Delta = 0.18$ (third row). For parameter sets in which the fully synchronized state was observed, its probability of occurrence is shown.
			}
			\label{Fig1}
		\end{figure}

		Fig.~\ref{Fig1} shows all evolutions of the order parameter grouped by several key sets of parameter values. For small amplitudes with a symmetric distribution about zero, bistability occurs between the fully synchronous state ($\lvert R \rvert = 1$) and chimera-like regimes ($\lvert R \rvert < 1$). For larger amplitudes bistability collapses as larger amplitudes of motion promote the loss of stability of the fully synchronous regime, as in the case considered in \cite{Smirnov2021}. When all amplitudes are non-negative, bistability does not occur, and we only obtain chimera-like regimes with oscillatory behavior of $\lvert R \rvert < 1$. For $\mu = 0.06$ the qualitative difference between $\lvert R \rvert$ for the cases of $\Delta = 0$ and $\Delta = 0.06$ seems negligible. However, as we show later, the exact microscopic structure differs between $\Delta = 0$ and $\Delta > 0$.  		
		
		We now describe the key dynamical regimes observed in the parameter space. The phase profiles $\varphi_n$ are displayed as a function of the particle positions $x_n$. We omit the case of bistability between the stationary chimera state and the fully synchronous state at $\mu = \Delta = 0$, when the particle grid is equidistant. We only plot the case of $\sigma = 0.1$, as for $\sigma = 0.03$ it is possible to obtain qualitatively the same regimes.

\begin{figure}
			\includegraphics[width = 1.0\linewidth]{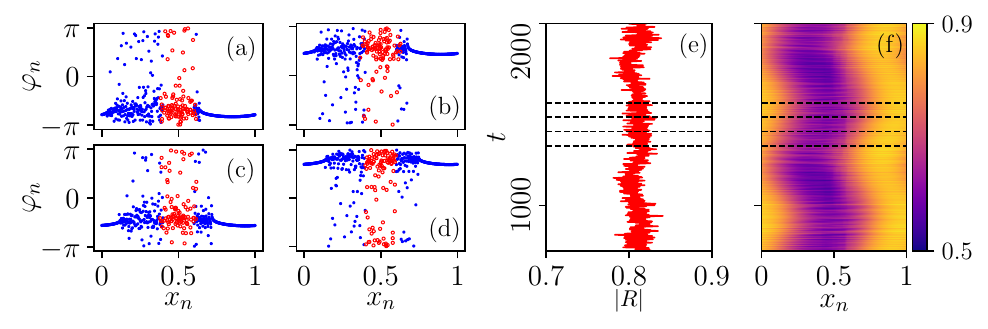}
			\caption{
				Periodically moving chimera state. (a--d) Phase snapshots at the time instants (a) $t = 1329$, (b) $t = 1407$, (c) $t = 1486$, (d) $t=1564$. Blue filled circular markers indicate stationary oscillators; red empty circular markers indicate moving oscillators. (e) Dynamics of $|R|$, black dotted horizontal lines indicate the time instants corresponding to panels (a--d). (f) Dynamics of $|H_n(t)|$. Parameters: $\sigma = 0.1, \mu = 0, \Delta = 0.06$.
			}
			\label{Fig2}
		\end{figure}

		In Fig.~\ref{Fig2}, where $\mu = 0$ and $\Delta = 0.06$, a periodically traveling chimera state is observed. The coherent cluster in Fig.~\ref{Fig2}(a--d, f) travels periodically. The global order parameter $|R(t)|$ in Fig.~\ref{Fig2}(e) exhibits oscillatory behavior, with fluctuations arising from finite-size effects ($N = 512$). Thus, the increase in the parameter $\Delta$ gave rise to a field perturbation that is spatially localized and periodic in time, thereby driving periodic dynamics of the phase profile across the medium. 
		
			\section{Atypical chimera states}
			\label{altern}
			
\begin{figure}
				\includegraphics[width = 1.0\linewidth]{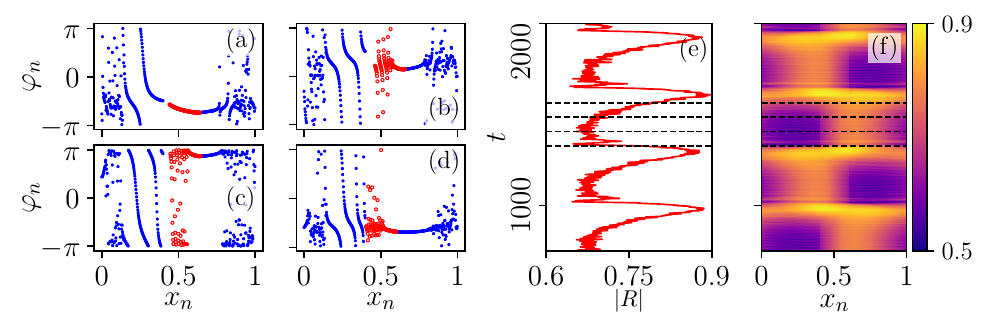}
				\caption{
					Alternating chimera state. Nonuniformly twisted state (from the perspective of microscopic dynamics). (a--f) The same as in Fig.~\ref{Fig2}. Parameters: $\sigma = 0.1, \mu = 0.06, \Delta = 0$.
				}
				\label{Fig4}
			\end{figure}

\begin{figure}
				\includegraphics[width = 1.0\linewidth]{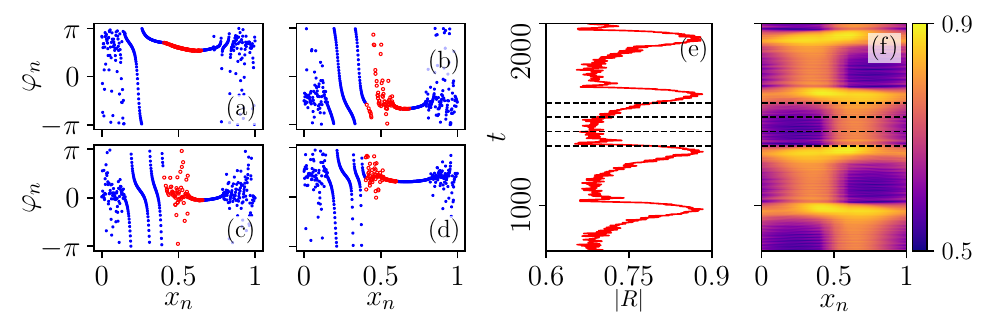}
				\caption{
					Alternating chimera state. Coherent-incoherent-twisted state (from the perspective of microscopic dynamics). (a--f) The same as in Fig.~\ref{Fig2}. Parameters: $\sigma = 0.1, \mu = 0.06, \Delta = 0.06$.
				}
				\label{Fig5}
			\end{figure}

			For an asymmetric amplitude distribution, alternating behavior is observed in Figs.~\ref{Fig4} and \ref{Fig5}. In both figures, $|R|$ exhibits relaxation-type oscillations and field $|H_n|$ demonstrates alternating behavior. The field exhibits long periods of stationary behavior, interrupted by brief, switching-like transitions that pass through states of nearly complete synchrony, after which the position of the coherent domain is shifted (panels (f)).

			\subsection{Field alternation induced by the effective kernel}
			\label{sec:4_1}

			To interpret the numerical results on alternating chimera states, we consider the dynamics of the field $|H_n|$ for small values of amplitudes corresponding to small displacements from the equidistant grid. 
			We approximate the interaction kernel $G(x_j(t)-x_n(t))$ by a Taylor polynomial of the first order around $x_j^{(0)}-x_n^{(0)}$ in terms of the displacement $ (A_j-A_n) \sin{\Omega t}$. This allows us to approximate the discrete mean field as
			\begin{equation*}
				H_n(t) \approx \frac{1}{N}\sum_{j=1}^N\bigl[ G(x_j^{(0)}-x_n^{(0)})+ (A_j-A_n)G'(x_j^{(0)}-x_n^{(0)})\sin{\Omega t} \bigr]\exp{i\varphi_j}.
			\end{equation*}
			It is convenient to consider the continuum limit, where the number of oscillators in any finite interval tends to infinity. In this case, we introduce the continuous mean field $H$, which is an average over infinitesimally close oscillators in the thermodynamic limit. Considering the motion amplitudes $A_k$ sufficiently small, we can assume approximate independence between phases $\varphi_j$ and amplitudes $A_j$ and average the amplitude factor $A_j - A_n$ independently of the phase factor $\exp{i\varphi_{j}}$,
			\begin{equation*}
				H(x, t) \approx \frac{1}{L}\int_0^L d\Tilde{x} \left\{ G(\Tilde{x}-x)+\left [\langle A \rangle (\Tilde{x})-\langle A \rangle (x) \right] G'(\Tilde{x}-x) \sin{\Omega t} \right \}\exp{i\varphi(\Tilde{x}, t)}
			\end{equation*}
			or
			\begin{equation*}
				H(x, t) \approx \frac{1}{L}\int_0^L \!\!\!\mathrm{d}\Tilde{x}\, G_{\mathrm{eff}}(x,\Tilde{x},t)\exp{i\varphi(\Tilde{x}, t)},
			\end{equation*}
			where
			\begin{equation}
				\label{eq:eff_kernel}
				G_{\mathrm{eff}}(x, \Tilde{x}, t) = G(\Tilde{x}-x)+\left [\langle A \rangle (\Tilde{x})-\langle A \rangle (x) \right] G'(\Tilde{x}-x) \sin{\Omega t}
			\end{equation}
			is the effective asymmetric kernel. 
			
\begin{figure}[t]
				\centering
				\includegraphics[width = 0.6\linewidth]{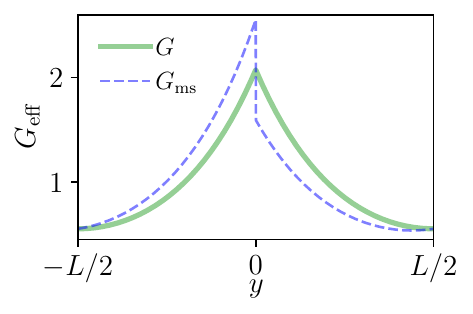}
				\caption{Effective interaction kernels between particles with positions $\Tilde{x}$ and $x$ at the time instants $t_k = (\pi/2 + 2\pi k)/\Omega$ ($k = 0,1,2,\ldots$). The argument $y = \Tilde{x}-x$. $G$ denotes the effective kernel for particles on an equidistant grid (green continuous curve), $G_{\mathrm{ms}}$ denotes the interaction kernel between a moving and a stationary particle (blue dashed line). Parameters: $\sigma = 0.1, \mu = 0.06$.}
				\label{Fig_eff_kernel}
			\end{figure}

			In our case, \begin{equation*}\langle A\rangle(x) = \begin{cases}
				\mu &\text{if } |x-L/2| \leq \sigma, \\
				0 & \text{otherwise.}
			\end{cases}\end{equation*}
			Thus, the effective interaction kernel between two stationary particles is simply $G(\Tilde{x}-x)$, while for a moving and a stationary particle, it is $G(\Tilde{x}-x) + \mu G'(\Tilde{x}-x) \sin{\Omega t}$. We plot these kernels in  Fig.~\ref{Fig_eff_kernel} at time instants when $\sin{\Omega t} = 1$. As seen, due to particle motion, the effective kernel—being the sum of an even and an odd function—becomes asymmetric, which explains both the formation of twisted regions and the shift of the coherent region, similarly to how it was demonstrated in~\cite{omelchenko2020travelling}.
			
		\subsection{Nonuniformly twisted state. Coherent-incoherent-twisted state}
		\label{sec:4_2}
		
\begin{figure}[t]
			\centering
			\includegraphics[width = 0.6\linewidth]{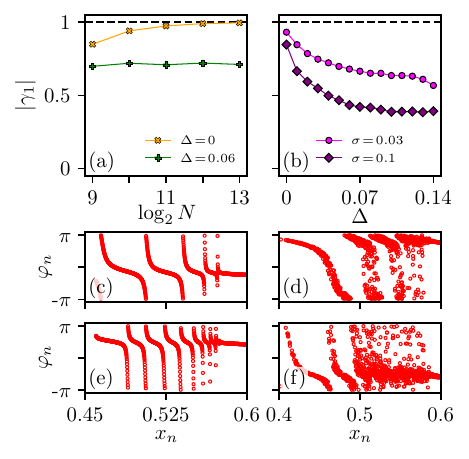}
			\caption{
				(a) Cross-correlation values $|\gamma_1|$, calculated via Eq.~\eqref{gamma_1} for $\sigma = 0.1$, $\mu = 0.06$, and $\Delta = 0$ or $0.06$ and averaged over the last period $T = 2\pi/\Omega$ of particle oscillation, as a function of the number of elements $N$. (b) Cross-correlation values $|\gamma_1|$ calculated via Eq.~\eqref{gamma_1} for $\sigma = 0.1$ or $0.03$, $\mu = 0.06$, $N = 512$ and averaged over the last period $T = 2\pi/\Omega$ of particle oscillation and additionally averaged over $30$ samples of moving particles amplitudes, for different values of distribution half-width $\Delta$. For $\Delta = 0$, the phases of moving particles are locally correlated; for $\Delta > 0$, only partial local phase correlation is observed among the moving particles. (c--f) Phase snapshots of the moving particles subgroup at the time instants $t = 1407$ (c,d) and $t = 1486$ (e,f): the fully correlated case (c,e) and the partially correlated case (d,f). Parameters: (c,e) $\sigma = 0.1, \mu = 0.06, \Delta = 0, N = 8192$, (d,f) $\sigma = 0.1, \mu = 0.06, \Delta = 0.06, N = 8192$. The initial conditions for (c) and (e) are the same, as well as for (d) and (f).
			}
			\label{Fig7}
		\end{figure}

		We analyze the microscopic phase dynamics of the twisted regions forming on the set of moving particles using cross-correlation analysis techniques similar to those employed in~\cite{smirnov2024nonuniformly}. We compute the cross-correlation of moving particle phases and average it over the last period of particle oscillation. First, moving particles are sorted by their coordinate $x_m$, resulting in a new index $\Tilde{m}$ for the $m$th particle, after which a M\"{o}bius transformation is applied to their phases:
		\begin{equation*}
			\exp{i\nu_{\Tilde{m}}} = (\exp{i\varphi_{\Tilde{m}}}-R_{M})/(1-R_{M}^*\exp{i\varphi_{\Tilde{m}}}),
		\end{equation*}
		where
		\begin{equation*}
			R_{M} = \frac{1}{ \lvert M\rvert} \sum_{\Tilde{m} \in M} \exp{i\varphi_{\Tilde{m}}}
		\end{equation*}
		is the order parameter for the subset of moving particles. We define the cross-correlation of the phase profile under a one-step circular shift as
		\begin{equation}
			\label{gamma_1}
			\gamma_1 = \frac{1}{ \lvert M\rvert} \sum_{\Tilde{m} \in M} \exp\bigl(i(\nu_{\Tilde{m}} - \nu_{\Tilde{m}+1 \bmod \lvert M\rvert})\bigr).
		\end{equation}
		Particles are assumed sufficiently dense for the shift to be nearly uniform. We consider only a single shift step, as our focus is on the presence of strong local correlation rather than on the full correlation profile. As seen in Fig.~\ref{Fig7}, the emergence of the incoherent cluster is due to $\Delta > 0$. In the case of in-phase motion with identical amplitudes ($\Delta = 0$), the twisted structures forming on the subset of moving particles are synchronous. On the other hand, when the oscillators move in phase with random amplitudes ($\Delta > 0$), partially synchronized twisted states emerge, driven by an asymmetry in the effective interaction kernel in Eq.~\eqref{eq:eff_kernel}.
		A detailed analysis of the cross-correlation dependence on $\Delta$ is presented in Fig.~\ref{Fig7}(b). As $\Delta$ increases, the correlation predictably drops.
		
		To understand the influence of local coherence properties of the moving elements subgroup on the whole system, we analyze two characteristic regimes at \(\mu = 0.06\): one with identical amplitudes (\(\Delta = 0\)) and one with finite amplitude heterogeneity (\(\Delta = 0.06\)).
		 For identical amplitudes ($\mu = 0.06$ and $\Delta = 0$), the system exhibits a nonuniformly twisted state, as seen in Fig.~\ref{Fig4}. Thus, an increase in $\mu$ leads to the formation of twisted regions due to the asymmetry of the effective kernel in Eq.~\eqref{eq:eff_kernel}.
		 This state lacks incoherent regions, as confirmed by numerical simulations. Disorder in the setup complicates the microscopic structure of the phase profile, inducing a localized region that is qualitatively similar to an incoherent state at finite $N = 512$.
		 
		 \begin{figure}
		 	\centering
		 	\includegraphics[width = 0.6\linewidth]{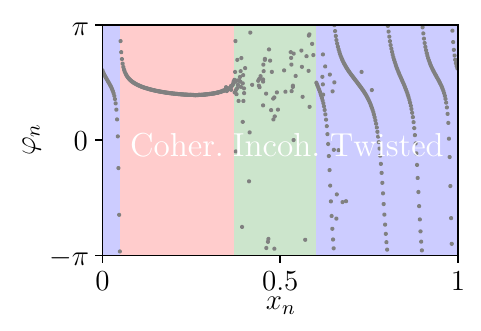}
		 	\caption{Coherent-incoherent-twisted state. Phase snapshot at the time instant $t = 1750$. The red region corresponds to the coherent domain, the green region to the incoherent domain, and the blue region to the twisted domain. Parameters: $\sigma = 0.1, \mu = 0.06, \Delta = 0.06$.}
		 	\label{Fig6}
		 \end{figure}
		 
		 Fig.~\ref{Fig5} shows the emergence of a coherent-incoherent-twisted state in the presence of non-zero amplitude heterogeneity ($\mu = 0.06$ and $\Delta = 0.06$). It features a purely incoherent region lacking local phase correlation among the moving particles. Fig.~\ref{Fig6} provides a phase snapshot at a representative time instant. The regime comprises three distinct regions: coherent, incoherent, and twisted. Note that the coherent and twisted regions switch periodically.
		 This state lacks incoherent regions among the stationary particles, as confirmed by numerical simulations.

		\section{Conclusion}
		\label{sec:Conclusion}
		
		In this work, we investigate the impact of periodic motion of elements within a localized spatial region on the phase dynamics of an ensemble of nonlocally coupled identical phase oscillators. In contrast to the case of equidistantly distributed oscillators—where the system exhibits bistability between a fully synchronized state and a chimera state—the periodic displacement of a subset of oscillators within a confined region can destabilize global synchronization and enrich the microscopic dynamics of the chimera regime.
		
		It was demonstrated that the motion of particle subgroups can induce asymmetry in the effective interaction kernel, resulting in the periodic drift of the chimera pattern. When the moving elements within the asynchronous domain of the chimera have equal amplitudes, a nonuniformly twisted state emerges, characterized by full phase correlation among neighboring oscillators. In contrast, for motion with random amplitudes, this correlation is disrupted, leading to asynchronous dynamics within the subset of moving particles. Furthermore, on the subset of stationary particles, a fully synchronous cluster coexists with the nonuniformly twisted state, collectively forming a coherent-incoherent-twisted state.
		

		\section*{Acknowledgments}
		
		This work was supported by the Ministry of Science and Higher Education of the Russian Federation under project No.~FSWR-2020-0036 (Sec.~\ref{fs_pm}) and by the Russian Science Foundation under projects No.~22-12-00348-P (Sec.~\ref{sec:4_1}) and No.~23-12-00180 (Sec.~\ref{sec:4_2}).

		
		
		\bibliographystyle{elsarticle-num} 
		\bibliography{literature.bib}

	\end{document}